\documentclass[10pt, a4paper, conference, compsocconf]{IEEEtran}

\ifCLASSINFOpdf
\else
\fi
\usepackage[cmex10]{amsmath}
\usepackage{multirow}
\usepackage{graphicx}
\usepackage{colortbl} 
\usepackage{tabularx}
\usepackage{color,soul}
\usepackage[table]{xcolor}
\usepackage{paralist}
\usepackage{longtable}
\usepackage{ccaption}
\usepackage{supertabular}
\usepackage{verbatim} 
\makeatletter

\begin{document}
%
\title{Combining Privileged Information to Improve Context-Aware Recommender Systems}

\author{\IEEEauthorblockN{Camila V. Sundermann\IEEEauthorrefmark{1},
Marcos A. Domingues\IEEEauthorrefmark{1},
Ricardo M. Marcacini\IEEEauthorrefmark{2} and
Solange O. Rezende\IEEEauthorrefmark{1}}
\\
\IEEEauthorblockA{\IEEEauthorrefmark{1}Instituto de Ci{\^e}ncias Matem{\'a}ticas e de Computa\c{c}{\~a}o\\
Universidade de S{\~a}o Paulo\\
S{\~a}o Carlos, SP, Brazil\\ $\lbrace$camilavs, mad, solange$\rbrace$@icmc.usp.br}
\\
\IEEEauthorblockA{\IEEEauthorrefmark{2}Universidade Federal do Mato Grosso do Sul\\
Tr{\^e}s Lagoas, MS, Brazil\\
ricardo.marcacini@ufms.br}}

\maketitle

\begin{abstract}

A recommender system is an information filtering technology which can be used to predict preference ratings of items (products, services, movies, etc) and/or to output a ranking of items that are likely to be of interest to the user. Context-aware recommender systems (CARS) learn and predict the tastes and preferences of users by incorporating available contextual information in the recommendation process. One of the major challenges in context-aware recommender systems research is the lack of automatic methods to obtain contextual information for these systems. Considering this scenario, in this paper, we propose to use contextual information from topic hierarchies of the items (web pages) to improve the performance of context-aware recommender systems. The topic hierarchies are constructed by an extension of the LUPI-based Incremental Hierarchical Clustering method that considers three types of information: traditional bag-of-words (technical information), and the combination of named entities (privileged information I) with domain terms (privileged information II). We evaluated the contextual information in four context-aware recommender systems. Different weights were assigned to each type of information. The empirical results demonstrated that topic hierarchies with the combination of the two kinds of privileged information can provide better recommendations. 
\end{abstract}



%
\IEEEpeerreviewmaketitle

\section{Introduction}
\label{sec:intro}

A recommender system is an information filtering technology which can be used to predict preference ratings of items (products, services, movies, etc) and/or to output a ranking of items that are likely to be of interest to the user \cite{Ricci2011}. This kind of system has emerged in order to reduce the difficulty of users to choose the product or service that most meets their needs. Many areas have been using recommender systems, mainly some web sites, like Amazon\footnote{http://www.amazon.com}, Netflix\footnote{http://www.netflix.com} and Last.fm\footnote{http://www.last.fm}.

Recommender systems usually use web access logs which represent the interaction activity between users and items. Traditional recommender systems consider only the two entities, items and users, to build the recommendation model. However, the use of contextual information can improve the recommendation process in some cases \cite{Adomavicius2005b, Domingues2013a}. The researchers that already investigated the use of context discovered that the quality of recommendations increases when additional information, like time, place, and so on, is used. 

The concept \textit{context} can assume different definitions. In this paper we consider that context is any information that can be used to characterize the situation of an entity \cite{Dey2001}. An example of application in which to consider contextual information can be important is movie recommendation. An user can prefer watch a love story with his girlfriend on Saturday night and a comedy with his friends during the week. So an online video store can recommend the movie that more corresponds to the user’s context. 

Although the proven importance of the use of contextual information in the recommendation process, there is still a lack of automatic methods to obtain such information. In a context-aware recommender system is possible to consider the context of the user or the context of the item. In this work we focus on the context extracted for the items (in our case, web pages).

The contextual information can be represented and structured in various ways. A form of organizing this information is using hierarchical structures. In \cite{Adomavicius2005b} and \cite{Panniello2012}, the researchers represented the context as trees. Given this possibility of hierarchical organization of context, we have been using topic hierarchies as a way to organize and extract the context of the textual content of web pages \cite{Domingues2014,Sundermann2014}. 

Most of the methods in the literature to build topic hierarchies represent the texts as a traditional bag-of-words, i.e., these methods consider the terms of texts as a disordered set of words. In \cite{Domingues2014}, we constructed topic hierarchies of web pages by using traditional bag-of-words, and the extracted topics were used as context of these pages in context-aware recommender systems. However, Marcacini and Rezende \cite{Marcacini2013} proposed a method, called LUPI-based Incremental Hierarchical Clustering (LIHC) to construct topic hierarchies that uses besides the bag-of-words (technical information), also the privileged information, which is a more valuable kind of information extracted from texts. In \cite{Sundermann2014}, we constructed topic hierarchies of the web pages using the method LIHC. We considered the bag-of-words and the named entities, extracted from the web pages, as privileged information, and we used the topics as the contextual information of the web pages in context-aware recommender systems. However, named entities are one of the various types of information that can be considered as privileged information.

The original LIHC method used only one type of privileged information to construct the topic hierarchies, in other words, it did not work with more than a kind of privileged information at the same time. In this paper we extend the method LIHC to be able to work with two kinds of privileged information, i.e., to construct topic hierarchies using besides the technical information, also other two kinds of information (privileged information). So, we propose to use topic hierarchies constructed by using three kinds of information: bag-of-words (technical information), named entities (privileged information I) and domain terms (privileged information II). The aim of this work is to combine this information and evaluate the impact of the use of the topics extracted from this combination as contextual information in context-aware recommender systems.  

This paper is structured as follows: in Section~\ref{sec:rw}, we report the related work. In Section~\ref{sec:proposal}, we present our proposal. We evaluate our proposal in Section~\ref{sec:evaluation}. And, finally, in Section~\ref{sec:conclusion}, we present conclusion and future work.

\section{Related Work}
\label{sec:rw}


There are three different ways to acquire contextual information: \textit{explicitly}, \textit{implicitly} and \textit{inferred} \cite{Adomavicius2011}. The explicit acquisition methods collect the contextual information through direct questions to the users. The implicit acquisition methods get contextual information directly from Web data or environment. The inference methods obtain contextual information using data an text mining techniques. In this paper, we infer context from web pages using text mining techniques. Following, some related works are presented.

In \cite{Li2010}, Li et al. proposed methods to extract contextual information from online reviews. They investigated available restaurant review data and four types of contextual information for a meal: the company (if the meal involved multiple people), occasion (for which occasions is the event), time (what time of the day) and location (in which city the event took place). They developed their algorithms by using existing natural language processing tools such as GATE tool\footnote{http://gate.ac.uk}. Hariri et al. \cite{Hariri2011} introduced a context-aware recommendation system that obtains contextual information by mining hotel reviews made by users, and combine them with user's rating historic to calculate a utility function over a set of items. They used a hotel review dataset from ``\textit{Trip Advisor website}''\footnote {http://www.tripadvisor.com}.

The methods proposed by Li et al. \cite{Li2010} and Hariri et al. \cite{Hariri2011} assume there are explicit contextual information in reviews, and such information is obtained for each review by mapping it to the labels. Therefore, they use supervised methods to learn the labels. The advantage of our proposal is that it exploits unsupervised methods to learn topic hierarchies. Therefore, it does not need a mapping between reviews and labels.

Aciar \cite{Aciar2010} proposed a technique to detect sentences of reviews with contextual information. She applied text mining tools to define sets of rules for identifying such sentences with context. In her work the phrases are classified into two categories: ``Contextual'' and ``Preferences''. The category ``Contextual'' groups phrases that present information on the context in which the review was written. The category ``Preferences'' groups phrases that present information about the features that consumers evaluated.

Our work differs from the Aciar's method since it is capable of using more text mining techniques, and these techniques are unsupervised, to extract contextual information. Aciar uses supervised techniques and conduct the evaluation of her method by using a case study, i.e., she does not compare her method results against other methods in the literature. Besides, she does not discuss the use of the extracted information in the recommendation process.

Ho et al. \cite{Ho2012} proposed an approach to mine future spatiotemporal events from news articles, and thus provide information for location-aware recommendation systems. A future event consists of its geographic location, temporal pattern, sentiment variable, news title, key phrase, and news article URL. Besides that, their method is unsupervised and also extracts topics.

In \cite{Ho2012}, the contextual information that Ho et al. extracted are related to time and local. The information of time is extracted from the timestamp of the article publication. To extract information of local, they also used named entity recognition. However, they did not evaluate the impact of the contextual information that they extracted in the recommender systems. The authors only presented some results about the evaluation of the context extraction process. 

Bauman and Tuzhilin \cite{Bauman2014} presented a method to find relevant contextual information from reviews of users. In this method, the reviews are classified as ``specific'' and ``generic''. They found that contextual information is contained mainly in the specific reviews, which are those that describe specific visits of a user to an establishment. Therefore, the context is extracted from the ``specific'' reviews by means of two methods: ``word-based'' and ``LDA-based''.

In \cite{Bauman2014}, Bauman and Tuzhilin consider that the contextual information is not known a priori. Besides that, their method is unsupervised and also extracts topics. Our method differs from theirs since it extracts topics using also privileged information, which enrich the contextual information.   

Our method has many advantages over the other ones proposed in the literature. In general, it does not need previous information (for example, labels). It uses unsupervised methods and combines technical information with privileged information, which enriches the contextual information. Additionally, the context extracted is about the item (web pages) and not the user. Finally, our results, presented in Section~\ref{sec:evaluation}, demonstrate that our contextual information is able to improve the quality of recommendations.

\section{Our Proposal}
\label{sec:proposal}

\begin{figure}[!htb] \centering
\centering{\includegraphics[width=6.5cm,height=2.84cm]{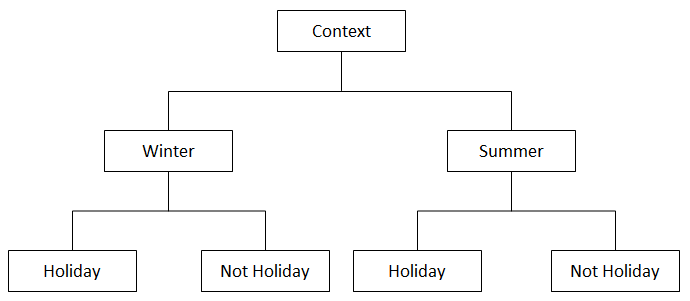}}
\caption{Hierarchical structure of the contextual attribute ``period of the year'' \cite{Panniello2012}.} \label{figcontexthie}\end{figure}

As already stated, the term context can assume many definitions depending on what area it is being treated in. We consider the definition given by Dey \cite{Dey2001} that says: ``Context is any definition that can be used to characterize the situation of an entity''. In our work the entities are web pages (items). Besides the definition, the contextual information can be represented using many structures. Some researchers treat the context as a hierarchical structure and represent it using trees. For example, Panniello and Gorgoglione \cite{Panniello2012} represent the attribute ``period of year'' as a tree like illustrated in Figure~\ref{figcontexthie}.

The idea of this research is representing the contextual information using a hierarchical structure called topic hierarchy. Topic hierarchies organize texts into groups and subgroups, and for each group, topics are extracted to represent the main issue of the group. Constructing a topic hierarchy of the items in a recommender system means grouping them by context, i.e., the topics extracted for each group represent the context of the group. Items of a same group are in the same context. We construct topic hierarchies by using the textual content of the web pages, and use the topics as contextual information in context-aware recommender systems.

Topics hierarchies can be constructed using hierarchical clustering. Traditional methods represent the textual collection as a bag-of-words \cite{Salton1983}, also known as technical information \cite{Vapnik2009}. However, we can extract concepts from the texts that are not represented in a simple bag-of-words. Named entities and domain terms are good examples of concepts that may be formed by a word or by more than a word and that are identified and extracted by using more advanced text preprocessing techniques. Thus, these two kinds of information, named entity and domain term, are considered in this paper as privileged information.

The term Named Entity was born in the Message Understanding Conferences (MUC) and includes names of people, organizations and locations, besides numeric expressions like time, date, money and percent expressions \cite{Sekine2004}. The named entity recognition is a task that involves identifying words or expressions that belong to categories of named entities \cite{Mikheev1999}. For example, in the sentence: ``Ana Maria works at Petrobras, in Brazil, since 1989''. ``Ana Maria'' is recognized as a person, ``Petrobras'' as an organization, ``Brazil'' as a location and ``1989'' as a date.

Despite the importance of term extraction task, there is still no consensus on the formal definition of what the ``term'' is. A definition widely accepted of term is given by Cabr{\'e} and Vivaldi \cite{Cabre2001}, which is: ``terminological unit obtained from specialized domain''. In most researches found in literature, the authors state that terms are generally nominal units, since they describe concepts. For example, in the Ecologic domain, the terms ``climate'', ``plant'', ``Atlantic forest'' and ``soil moisture'' are examples of domain terms \cite{Conrado2014}. The terms are used in applications such as information retrieval, information extraction and summarization.

In our proposal, we instantiate the LUPI-based Incremental Hierarchical Clustering (LIHC) method \cite{Marcacini2013} to construct topic hierarchies using one type of privileged information and technical information. Let $D^{pri} = \lbrace d_{1}^p,\dots, d_{m}^p \rbrace$ and $D^{tec} = \lbrace d_{1}^t,\dots, d_{m}^t, d_{m+1}^t,\dots, d_n \rbrace$ the sets of documents represented by the privileged information (totaling $m$ documents) and with technical information (totaling $n$ documents), respectively, where $d^p \in D^p$ and $d^t \in D^t$. Note that the number of documents represented by the privileged information, in general, is smaller than the number of documents represented by technical information, i.e, $m \leq n$. This is due to the fact that a significant number of documents do not contain features extracted from privileged information (\textit{e.g.}, named entities and domain terms).

The subset of documents that contain  the privileged information and technical information, $Y = \lbrace (d_{1}^t, d_{1}^p),\dots, (d_{m}^t, d_{m}^p) \rbrace$, is used for learning the initial clustering model. In this case, various clustering algorithms are run (or repeated runs of the same algorithm with different parameter values) to obtain several clusters from the subset $Y$. To aggregate the generated clusters, the LIHC method obtains two co-association matrices $M^{t}(i,j)$ and $M^{p}(i,j)$ which represent, respectively, the technical information (bag-of-words) clustering model and privileged information clustering model. The combination of these two clustering models is performed by using a consensual co-association matrix:
\begin{equation}
M^F(i,j)=  (1-\alpha) M^t(i,j) + \alpha M^p(i,j),
\end{equation} for all items $i$ and $j$. In this case, the parameter $ \alpha $ is a combination factor $ (0 \leq \alpha \leq 1) $ that indicates the importance of the privileged information space in the final co-association matrix. The initial model of the LIHC method is obtained by applying any hierarchical clustering algorithm from the matrix $M^F$. The remaining text documents, i.e., the documents without privileged information, are inserted incrementally into hierarchical clustering by using the nearest neighbor technique. For the construction of topic hierarchies, the topic extraction is based on selection of the most frequent terms of each cluster. 

In \cite{Sundermann2014}, we constructed topic hierarchies of the web pages by using the method
LIHC and considering as privileged information only named entities. In this paper, we construct topic hierarchies by combining named entities and domain terms as privileged information, varying the weight of each type of information. To incorporate the two types of privileged information, the method LIHC was extended as follows.



First, $D^{pri}$ is divided into two sets $D^{ne}$ (the set of the privileged information I – named entities) and $D^{dt}$ (the set of the privileged information II – domain terms). Let $D^{ne}=\lbrace d_1^{ne},\dots,d_r^{ne} \rbrace$  be the set of documents with named entities (totaling $r$ documents) and $D^{dt}=\lbrace d_1^{dt},\dots,d_s^{dt} \rbrace$ be the set of documents with domain terms (totaling $s$ documents). Similarly, the matrix $M^p(i,j)$ is divided into two matrices $M^{ne}$ (the named entities clustering model) and $M^{dt}$ (the domain terms clustering model). The combination of the three clustering models ($M^t$, $M^{ne}$ and $M^{dt}$) is performed by using the follow consensual co-association matrix:
\begin{equation}
M^{nf}(i,j) = (1-\alpha)M^t(i,j) + \beta M^{ne}(i,j) + \theta M^{dt}(i,j),
\end{equation} where, $\beta$ and $\theta$ indicate  the importance of the named entities and domain terms, respectively,  in the final co-association matrix, and $\beta + \theta = \alpha$. In the next section, we empirically evaluate our proposal by using different values of $\alpha$,  $\beta$ and $\theta$ to construct the topic hierarchies.

\section{Empirical Evaluation}
\label{sec:evaluation}

The aim of our work is to study the impact of the context, extracted by our method, in context-aware recommender systems. So, the empirical evaluation consists of comparing the results of the algorithms \textit{C. Reduction} \cite{Adomavicius2005b}, \textit{DaVI-BEST} \cite{Domingues2013a}, \textit{Weight PoF} and \textit{Filter PoF} \cite{Panniello2012}, all them using our contextual information, against the uncontextual algorithm \textit{Item-Based Collaborative Filtering} (IBCF) \cite{Deshpande2004}. In this way, we compared the quality of the recommendations generated by using our context against the quality of the recommendations generated without using contextual information. In this section, we present the necessary details to understand our experiments: data set, baseline, context-aware recommender algorithms, experimental setup, evaluation measure and the results. 

\subsection{Data Set}
\label{sub:data}

In the experiments we used a data set from a Portuguese website about agribusiness that consists of 4,659 users, 15,037 accesses and 1,543 web pages written in Portuguese language. To construct the topic hierarchies for these web pages, we used the textual content of the pages, eliminating header, footer and everything that do not pertaining to the main textual content.

We preprocessed the texts executing traditional text preprocessing tasks: stopword removal and stemming. The representations or ``term value matrix'' were constructed using the term weighting measure TF-IDF (term frequency-inverse document frequency). Three representations were constructed: the traditional bag-of-words representation (technical information), the named entities representation (privileged information I) and the domain terms representation (privileged information II). We defined the weights testing different combinations of the two kinds of privileged information. The weights are shown in Table~\ref{tab:weights}. 


\begin{table}[htb] \centering
\caption{Values of combination factor $\alpha$ used in this paper and the weights of each information}\label{tab:weights}
\begin{tabular}{|c|c|c|c|}
\hline 
\multirow{2}{*}{Combination Factor} & \multirow{2}{*}{Technical Information} & \multicolumn{2}{c|}{Privileged Information}\\ 
\cline{3-4}
 & & Named Entities & Domain Terms\\
\cline{1-2}\cline{2-3} \cline{3-4}
\multirow{3}{*}{$\alpha = 0.3$} & \multirow{3}{*}{$70\%$} & $10\%$ & $20\%$ \\ 
\cline{3-4}
 & & $20\%$ & $10\%$\\
\cline{3-4}
 & & $15\%$ & $15\%$\\
\cline{1-2}\cline{2-3} \cline{3-4} 
\multirow{3}{*}{$\alpha = 0.5$} & \multirow{3}{*}{$50\%$} & $20\%$ & $30\%$ \\ 
\cline{3-4}
 & & $30\%$ & $20\%$\\
\cline{3-4}
 & & $25\%$ & $25\%$\\
\cline{1-2}\cline{2-3} \cline{3-4}
\multirow{3}{*}{$\alpha = 0.7$} & \multirow{3}{*}{$30\%$} & $20\%$ & $50\%$ \\ 
\cline{3-4}
 & & $50\%$ & $20\%$\\
\cline{3-4}
 & & $35\%$ & $35\%$\\
\cline{1-2}\cline{2-3} \cline{3-4}
\multirow{3}{*}{$\alpha = 1$} & \multirow{3}{*}{$0\%$} & $30\%$ & $70\%$ \\ 
\cline{3-4}
 & & $70\%$ & $30\%$\\
\cline{3-4}
 & & $50\%$ & $50\%$\\
\hline 
\end{tabular}
\end{table}

We extracted the topics from the topic hierarchies considering three configurations $\lbrace 50,100 \rbrace$,  $\lbrace 15,20 \rbrace$ and $\lbrace 2,7 \rbrace$. In this configuration $\lbrace x,y \rbrace$, that represents the granularity level, the parameter $x$ identifies the minimum number of items allowed in the topic, while the parameter $y$ identifies the maximum number of items per topic. Topics with more items associated to them mean topics more generic, while topics with fewer items associated to them mean topics more specific. So, the topics extracted by the configuration $\lbrace 50,100 \rbrace$ are more generic and the topics extracted by the configuration $\lbrace 2,7 \rbrace$ are more specific. The configuration $\lbrace 15,20 \rbrace$ are between the two others, and it was chosen because, in previous experiments, we obtained good results using this granularity level. Besides that, the more generic configuration extracts a lower number of topics, while the more specific configuration extracts a higher number of topics. Therefore, using this configuration we can analyses if the number of topics extracted or the granularity level of this topics influences the quality of the recommendations. In Table~\ref{tab:topics}, we can see the number of topics extracted using each configuration.

\begin{table}[htb] \centering
\caption{Extracted Topics}\label{tab:topics}
\begin{tabular}{|c|c|c|c|c|}
\hline 
\multicolumn{2}{|c|}{Weights} & \multicolumn{3}{c|}{Granularities} \\ 
\hline
Named Entities & Domain Terms & $\lbrace 50,100 \rbrace$ & $\lbrace 15,20 \rbrace$ & $\lbrace 2,7 \rbrace$\\
\hline
 $10\%$ & $20\%$ & 46 & 66 & 897\\ 
\hline
$20\%$ & $10\%$ & 42 & 61 & 902\\
\hline
$15\%$ & $15\%$ & 44 & 61 & 890\\
\hline
$20\%$ & $30\%$ & 46 & 66 & 869 \\ 
\hline
$30\%$ & $20\%$ & 49 & 73 & 872\\
\hline
$25\%$ & $25\%$ & 51 & 72 & 868\\
\hline
$20\%$ & $50\%$ & 45 & 61 & 892 \\ 
\hline
$50\%$ & $20\%$ & 46 & 70 & 901\\
\hline
$35\%$ & $35\%$ & 42 & 63 & 879 \\
\hline
$30\%$ & $70\%$ & 39 & 60 & 915\\ 
\hline
$70\%$ & $30\%$  & 35 & 69 & 870\\
\hline
$50\%$ & $50\%$ & 46 & 71 & 921\\
\hline 
\end{tabular}
\end{table}

\subsection{Supporting Tools and Methods}

In the experiments we used JPretext\footnote{http://sites.labic.icmc.usp.br/torch/msd2011/jpretext} and LIHC\footnote{ http://sites.labic.icmc.usp.br/torch/doceng2013} for the pre-processing and the hierarchical clustering of the items. These two tools are part of Torch \cite{Marcacini2010a}, that is a set of tools developed to support text clustering and construction of topic hierarchies. JPretext transforms the collection of texts in a ``term value matrix'' and LIHC tool implements the LUPI-based Incremental Hierarchical Clustering method.

The named entity recognition was performed by using REMBRANDT \cite{Cardoso2008}, a system that recognizes classes of named entities, like things, location, organization, people and others, in texts written in Portuguese. REMBRANDT uses Wikipedia\footnote {https://www.wikipedia.org} as knowledge base for the classification of the entities.
 
Lastly, for the domain term extraction we used the method MATE-ML (Automatic Term Extraction based on Machine Learning) \cite{Conrado2013, Conrado2014}. This method uses machine learning incorporating rich features of candidate terms. The steps of MATE-ML are: 1) Text Pre-Processing; 2) Extraction of linguistic, statistic and hybrid features; 3) Application of filters; and 4) Generation of inductive models based on machine learning.

\subsection{Baseline}

In this paper we considered the un-contextual algorithm \textit{Item-Based Collaborative Filtering} (IBCF) \cite{Deshpande2004} as baseline. Let $m$ be the number of users $U = \lbrace u_1, u_2, ..., u_m \rbrace$ and $n$ the number of items that can be recommended $I = \lbrace i_1, i_2, ..., i_n \rbrace$. An item-based collaborative filtering model $M$ is a matrix representing the similarities among all pairs of items, according to a similarity measure. We used the cosine angle similarity measure, defined as: \begin{equation}sim(i_1, i_2) = cos(\overrightarrow{i_1}, \overrightarrow{i_2}) = \dfrac{\overrightarrow{i_1} \cdot \overrightarrow{i_2}}{\Vert \overrightarrow{i_1} \Vert \ast \Vert \overrightarrow{i_2} \Vert},
\end{equation}where $\overrightarrow{i_1}$ and $\overrightarrow{i_2}$ are rating vectors and the operator ``$\cdot$'' denotes the dot-product of the two vectors. In our case, as we are dealing only with implicit feedback, the rating vectors are binary. The value 1 means that the user accessed the respective item, whereas the value 0 is the opposite.

Given an active user $u_a$ and his set of observable items $O \subseteq I$, the $N$ recommendations are generated as follows. First, we identify the set of candidate items for recommendation $R$ by selecting from the model all items $i \notin O$. Then, for each candidate item $r \in R$, we calculate its recommendation score as:\begin{equation}
score(u_a, O, r) = \dfrac{\sum_{i \in K_r \cap O}sim(r, i)}{\sum_{i \in K_r}sim(r, i)},
\end{equation}
where $K_r$ is the set of the $k$ most similar items to the candidate item $r$. The $ N $ candidate items with the highest values of score are recommended to the user $u_a$. 

All the context-aware recommendation algorithms used in this work are based on the Item-Based Collaborative Filtering. They are presented in the next section.

\subsection{Context-Aware Recommender Systems}
\label{sec:cars}

Context-aware recommender systems (CARS) learn and predict the tastes and preferences of users by incorporating available contextual information in the recommendation process. According to Adomavicius and Tuzhilin \cite{Adomavicius2011}, contextual information can be applied at various stages of the recommendation process. Following this criterion, these systems can be divided into three categories: contextual pre-filtering, contextual modeling and contextual post-filtering.

In this work, we evaluate the effects of using the contextual information, obtained from topic hierarchies, in four different context-aware recommender systems:

\begin{itemize}

\item \textbf{\textit{C. Reduction}} \cite{Adomavicius2005b} (Pre-filtering approach): in pre-filtering approaches the contextual information is used as a label for filtering out those data that do not correspond to the specified contextual information. The remaining data that passed the filter (contextualized data) is used to generate the model. \textit{C. Reduction} uses the contextual information as label to segment the data. A recommendation method is run for each contextual segment to determine which segment outperforms the traditional un-contextual recommendation model. The best contextual model is chosen to make the recommendation. Here the best model is the one that has the highest F1 measure.    

\item \textbf{\textit{DaVI-BEST}} \cite{Domingues2013a} (Contextual modeling approach): in this approach the context is used in the recommendation model, i.e., the contextual information is part of the model together with user and item data. \textit{DaVI-BEST} considers the contextual information as virtual items, using them along with the actual items in the recommendation model. All contextual information are evaluated and it is selected the dimension which better outperforms the traditional un-contextual recommendation model to make contextual recommendations.

\item \textbf{\textit{Weight PoF} and \textit{Filter PoF}} \cite{Panniello2012} (Contextual post-filtering approaches): these approaches use the contextual information to reorder and filter out the recommendation, respectively. Firstly, they apply the traditional algorithm to build the un-contextual recommendation model, ignoring the contextual information. Then, the probability of users to access the items given the right context is calculated. This probability is multiplied by scores of items to reorder the recommendations (\textit{Weight PoF}) or is used as a threshold to filter them (\textit{Filter PoF}).

\end{itemize}

\subsection{Experimental Setup and Evaluation Measures}
\label{sub:expe}

The protocol considered in this paper to measure the predictive ability of the recommender systems is the All But One protocol \cite{Breese1998} with 10-fold cross validation, i.e., the set of documents is partitioned into 10 subsets. For each fold we use $n-1$ of these subsets for training and the rest for testing. The training set $T_r$ is used to build the recommendation model. For each user in the test set $T_e$, an item is hidden as a singleton set $H$. The remaining items represent the set of observable items $O$, that is used in the recommendation. Then, we compute Mean Average Precision ($MAP@N$), where $N$ equals 5 and 10 recommendations. For each configuration and measure, the 10-fold values are summarized by using mean and standard deviation. To compare two recommendation algorithms, we applied the two-sided paired t-test with a 95$\%$ confidence level.

In our empirical evaluation, we used the 4 most similar items to make the recommendations and 0.1 as a threshold in \textit{Filter PoF} to filter out the recommendations, since these values provided the best results for this experiment.

\subsection{Results}
\label{sub:results}

In Table~\ref{tab:MAP1}, we show the results of our ranking evaluation by means of $MAP@N$. The results are obtained at four values of the combination factor ($\alpha = 0.3$, $\alpha = 0.5$, $\alpha = 0.7$ and $\alpha = 1$) and at three granularity levels, as described in Section~\ref{sub:data}. For each value of combination factor we also have the weights of each type of privileged information. To facilitate the understanding of the results, we mentioned the weight of technical information as BOW, the weight of the named entities as NE and the weight of domain terms as DT.  

\begin{table*}[!htpb] \centering
\caption{Comparing the context-aware recommendation algorithms against the IBCF algorithm. The values that are not statistically better than baseline ($p$-value$>$0.05) are in colored cells and the best results (higher values of MAP) for each algorithm are in boldface}\label{tab:MAP1}
\begin{tabular}{|c|c|c|c|c|c|c|c|c|}
\hline 
\multicolumn{9}{|c|}{MAP results for $\alpha = 0.3$ ($BOW=70\%$ - $EN=10\%$ - $DT=20\%$)}\\
\hline
\multirow{2}{*}{Granularities} & \multicolumn{4}{c|}{$MAP@5$} & \multicolumn{4}{c|}{$MAP@10$}\\ 
\cline{2-3}\cline{3-4}\cline{4-5}\cline{5-6}\cline{6-7}\cline{7-8}\cline{8-9}
& IBCF & \textit{C. Reduction} & \textit{Weight PoF} & \textit{Filter PoF} & IBCF & \textit{C. Reduction} & \textit{Weight PoF} & \textit{Filter PoF}\\
\hline 
$\lbrace 50,100 \rbrace$ & 0.2991455 & 0.3132490 & 0.3297517 & \cellcolor{blue!25} 0.1914374 & 0.3089337 & 0.3227017 & 0.3386906 & \cellcolor{blue!25} 0.1930766 \\
\hline
$\lbrace 15,20 \rbrace$ & 0.2991455 & \pmb{0.4190486} & \pmb{0.4615741} & 0.3435903 & 0.3089337 & \pmb{0.4228415} & \pmb{0.4651130} & 0.3446704 \\
\hline 
$\lbrace 2,7 \rbrace$ & 0.2991455 & 0.3315996 & 0.3992454 & \pmb{0.4267015} & 0.3089337 & 0.3397160 & 0.4056149 & \pmb{0.4278348} \\
\hline
\multicolumn{9}{|c|}{MAP results for $\alpha = 0.3$ ($BOW=70\%$ - $EN=20\%$ - $DT=10\%$)}\\
\hline
\multirow{2}{*}{Granularities} & \multicolumn{4}{c|}{$MAP@5$} & \multicolumn{4}{c|}{$MAP@10$}\\ 
\cline{2-3}\cline{3-4}\cline{4-5}\cline{5-6}\cline{6-7}\cline{7-8}\cline{8-9}
& IBCF & \textit{C. Reduction}& \textit{Weight PoF} & \textit{Filter PoF} & IBCF & \textit{C. Reduction} & \textit{Weight PoF} & \textit{Filter PoF}\\ 
\hline 
$\lbrace 50,100 \rbrace$ & 0.2991455 & 0.3281412 & 0.3464556 & \cellcolor{blue!25} 0.1823619 & 0.3089337 & 0.3382604 & 0.3569279 & \cellcolor{blue!25} 0.1842473 \\
\hline 
$\lbrace 15,20 \rbrace$ & 0.2991455 & \pmb{0.4242335} & \pmb{0.4496521} & 0.3210554 & 0.3089337 & \pmb{0.4289756} & \pmb{0.4537685} & \cellcolor{blue!25} 0.3231635 \\
\hline 
$\lbrace 2,7 \rbrace$ & 0.2991455 & 0.3220141 & 0.3754638 & \pmb{0.4181398} & 0.3089337 & 0.3298058 & 0.3815615 & \pmb{0.4196089}\\
\hline
\multicolumn{9}{|c|}{MAP results for $\alpha = 0.3$ ($BOW=70\%$ - $EN=15\%$ - $DT=15\%$)}\\
\hline
\multirow{2}{*}{Granularities} & \multicolumn{4}{c|}{$MAP@5$} & \multicolumn{4}{c|}{$MAP@10$}\\ 
\cline{2-3}\cline{3-4}\cline{4-5}\cline{5-6}\cline{6-7}\cline{7-8}\cline{8-9}
& IBCF & \textit{C. Reduction} & \textit{Weight PoF} & \textit{Filter PoF} & IBCF & \textit{C. Reduction} & \textit{Weight PoF} & \textit{Filter PoF}\\ 
\hline 
$\lbrace 50,100 \rbrace$ & 0.2991455 & 0.3338906 & 0.3478617 & \cellcolor{blue!25} 0.1865615 & 0.3089337 & 0.3437439 & 0.3571797 & \cellcolor{blue!25} 0.1876844 \\
\hline 
$\lbrace 15,20 \rbrace$ & 0.2991455 & \pmb{0.3702801} & \pmb{0.4175135} & \cellcolor{blue!25} 0.3040427 & 0.3089337 & \pmb{0.3762570} & \pmb{0.4227551} & \cellcolor{blue!25} 0.3053225 \\
\hline 
$\lbrace 2,7 \rbrace$ & 0.2991455 & 0.3260990 & 0.3891643 & \pmb{0.4063760} & 0.3089337 & 0.3356745 & 0.3960491 & \pmb{0.4077946} \\
\hline \hline
\multicolumn{9}{|c|}{MAP results for $\alpha = 0.5$ ($BOW=50\%$ - $EN=20\%$ - $DT=30\%$)}\\
\hline
\multirow{2}{*}{Granularities} & \multicolumn{4}{c|}{$MAP@5$} & \multicolumn{4}{c|}{$MAP@10$}\\ 
\cline{2-3}\cline{3-4}\cline{4-5}\cline{5-6}\cline{6-7}\cline{7-8}\cline{8-9}
& IBCF & \textit{C. Reduction} & \textit{Weight PoF} & \textit{Filter PoF} & IBCF & \textit{C. Reduction} & \textit{Weight PoF} & \textit{Filter PoF}\\ 
\hline 
$\lbrace 50,100 \rbrace$ & 0.2991455 & 0.3482771 & 0.3792558 & \cellcolor{blue!25} 0.2252132 & 0.3089337 & 0.3583244 & 0.3886332 & \cellcolor{blue!25} 0.2267441 \\
\hline 
$\lbrace 15,20 \rbrace$ & 0.2991455 & \pmb{0.3895239} & \pmb{0.4189020} & \cellcolor{blue!25} 0.2713655 & 0.3089337 & \pmb{0.3952202} & \pmb{0.4239535} & \cellcolor{blue!25} 0.2720121 \\
\hline 
$\lbrace 2,7 \rbrace$ & 0.2991455 & 0.3409991 & 0.3962494 & \pmb{0.4187704} & 0.3089337 & 0.3504907 & 0.4038026 & \pmb{0.4200874} \\
\hline
\multicolumn{9}{|c|}{MAP results for $\alpha = 0.5$ ($BOW=50\%$ - $EN=30\%$ - $DT=20\%$)}\\
\hline
\multirow{2}{*}{Granularities} & \multicolumn{4}{c|}{$MAP@5$} & \multicolumn{4}{c|}{$MAP@10$}\\ 
\cline{2-3}\cline{3-4}\cline{4-5}\cline{5-6}\cline{6-7}\cline{7-8}\cline{8-9}
& IBCF & \textit{C. Reduction} & \textit{Weight PoF} & \textit{Filter PoF} & IBCF & \textit{C. Reduction} & \textit{Weight PoF} & \textit{Filter PoF}\\ 
\hline 
$\lbrace 50,100 \rbrace$ & 0.2991455 & 0.3468419 & 0.3675248 & \cellcolor{blue!25} 0.1904683 & 0.3089337 & 0.3567991 & 0.3774468 & \cellcolor{blue!25} 0.1912013 \\
\hline 
$\lbrace 15,20 \rbrace$ & 0.2991455 & \pmb{0.3854856} & \pmb{0.4314550} & \cellcolor{blue!25} 0.3205283 & 0.3089337 & \pmb{0.3916562} & \pmb{0.4366166} & \cellcolor{blue!25} 0.3208468 \\
\hline 
$\lbrace 2,7 \rbrace$ & 0.2991455 & \cellcolor{blue!25} 0.3250967 & 0.3715466 & \pmb{0.4021174} & 0.3089337 & \cellcolor{blue!25} 0.3342467 & 0.3790192 & \pmb{0.4037792} \\
\hline
\multicolumn{9}{|c|}{MAP results for $\alpha = 0.5$ ($BOW=50\%$ - $EN=25\%$ - $DT=25\%$)}\\
\hline
\multirow{2}{*}{Granularities} & \multicolumn{4}{c|}{$MAP@5$} & \multicolumn{4}{c|}{$MAP@10$}\\ 
\cline{2-3}\cline{3-4}\cline{4-5}\cline{5-6}\cline{6-7}\cline{7-8}\cline{8-9}
& IBCF & \textit{C. Reduction} & \textit{Weight PoF} & \textit{Filter PoF} & IBCF & \textit{C. Reduction} & \textit{Weight PoF} & \textit{Filter PoF}\\ 
\hline
$\lbrace 50,100 \rbrace$ & 0.2991455 & 0.3228372 & 0.3427176 & \cellcolor{blue!25} 0.1882432 & 0.3089337 & 0.3318865 & 0.3514913 & \cellcolor{blue!25} 0.1896786 \\
\hline 
$\lbrace 15,20 \rbrace$ & 0.2991455 & \pmb{0.3742469} & \pmb{0.4267049} & \cellcolor{blue!25} 0.3011386 & 0.3089337 & \pmb{0.3795265} & \pmb{0.4310578} & \cellcolor{blue!25} 0.3023229 \\
\hline 
$\lbrace 2,7 \rbrace$ & 0.2991455 & 0.3399507 & 0.3897151 & \pmb{0.4343196} & 0.3089337 & 0.3497889 & 0.3975618 & \pmb{0.4358158} \\
\hline
\end{tabular}
\end{table*}

\begin{table*}[!htpb] \centering
\contcaption{(Continued) Comparing the context-aware recommendation algorithms against the IBCF algorithm. The values that are not statistically better than baseline ($p$-value$>$0.05) are in colored cells and the best results (higher values of MAP) for each algorithm are in boldface}
\begin{tabular}{|c|c|c|c|c|c|c|c|c|}
\hline 
\multicolumn{9}{|c|}{MAP results for $\alpha = 0.7$ ($BOW=30\%$ - $EN=20\%$ - $DT=50\%$)}\\
\hline
\multirow{2}{*}{Granularities} & \multicolumn{4}{c|}{$MAP@5$} & \multicolumn{4}{c|}{$MAP@10$}\\ 
\cline{2-3}\cline{3-4}\cline{4-5}\cline{5-6}\cline{6-7}\cline{7-8}\cline{8-9}
& IBCF & \textit{C. Reduction} & \textit{Weight PoF} & \textit{Filter PoF} & IBCF & \textit{C. Reduction} & \textit{Weight PoF} & \textit{Filter PoF}\\ 
\hline
$\lbrace 50,100 \rbrace$ & 0.2991455 & 0.3253358 & 0.3472713 & \cellcolor{blue!25} 0.2017587 & 0.3089337 & 0.3355101 & 0.3572727 & \cellcolor{blue!25} 0.2033553 \\
\hline 
$\lbrace 15,20 \rbrace$ & 0.2991455 & \pmb{0.3694954} & \pmb{0.4067977} & \cellcolor{blue!25} 0.2962737 & 0.3089337 & \pmb{0.3763551} & \pmb{0.4126285} & \cellcolor{blue!25} 0.2969542 \\
\hline 
$\lbrace 2,7 \rbrace$ & 0.2991455 & 0.3340217 & 0.3949698 & \pmb{0.4194938} & 0.3089337 & 0.3433043 & 0.4022932 & \pmb{0.4207608} \\
\hline
\multicolumn{9}{|c|}{MAP results for $\alpha = 0.7$ ($BOW=30\%$ - $EN=50\%$ - $DT=20\%$)}\\
\hline
\multirow{2}{*}{Granularities} & \multicolumn{4}{c|}{$MAP@5$} & \multicolumn{4}{c|}{$MAP@10$}\\ 
\cline{2-3}\cline{3-4}\cline{4-5}\cline{5-6}\cline{6-7}\cline{7-8}\cline{8-9}
& IBCF & \textit{C. Reduction} & \textit{Weight PoF} & \textit{Filter PoF} & IBCF & \textit{C. Reduction} & \textit{Weight PoF} & \textit{Filter PoF}\\ 
\hline
$\lbrace 50,100 \rbrace$ & 0.2991455 & 0.3364125 & 0.3512662 & \cellcolor{blue!25} 0.1996443 & 0.3089337 & 0.3445030 & 0.3599979 & \cellcolor{blue!25} 0.2008481 \\
\hline 
$\lbrace 15,20 \rbrace$ & 0.2991455 & \pmb{0.3606417} & \pmb{0.3957279} & \cellcolor{blue!25} 0.2848784 & 0.3089337 & \pmb{0.3668573} & \pmb{0.4012059} & \cellcolor{blue!25} 0.2859318 \\
\hline 
$\lbrace 2,7 \rbrace$ & 0.2991455 & 0.3303324 & 0.3840207 & \pmb{0.4041100} & 0.3089337 & 0.3393601 & 0.3918221 & \pmb{0.4059656} \\
\hline
\multicolumn{9}{|c|}{MAP results for $\alpha = 0.7$ ($BOW=30\%$ - $EN=35\%$ - $DT=35\%$)}\\
\hline
\multirow{2}{*}{Granularities} & \multicolumn{4}{c|}{$MAP@5$} & \multicolumn{4}{c|}{$MAP@10$}\\ 
\cline{2-3}\cline{3-4}\cline{4-5}\cline{5-6}\cline{6-7}\cline{7-8}\cline{8-9}
& IBCF & \textit{C. Reduction} & \textit{Weight PoF} & \textit{Filter PoF} & IBCF & \textit{C. Reduction} & \textit{Weight PoF} & \textit{Filter PoF}\\ 
\hline
$\lbrace 50,100 \rbrace$ & 0.2991455 & 0.3269617 & 0.3501964 & \cellcolor{blue!25} 0.2149991 & 0.3089337 & 0.3359110 & 0.3587213 & \cellcolor{blue!25} 0.2163066 \\
\hline 
$\lbrace 15,20 \rbrace$ & 0.2991455 & \pmb{0.3809494} & \pmb{0.4162618} & \cellcolor{blue!25} 0.2924863 & 0.3089337 & \pmb{0.3880923} & \pmb{0.4215766} & \cellcolor{blue!25} 0.2940350 \\
\hline 
$\lbrace 2,7 \rbrace$ & 0.2991455 & 0.33008407 & 0.3867889 & \pmb{0.4177925} & 0.3089337 & 0.3400565 & 0.3944094 & \pmb{0.4188557} \\
\hline \hline
\multicolumn{9}{|c|}{MAP results for $\alpha = 1$ ($BOW=0$ - $EN=30\%$ - $DT=70\%$)}\\
\hline
\multirow{2}{*}{Granularities} & \multicolumn{4}{c|}{$MAP@5$} & \multicolumn{4}{c|}{$MAP@10$}\\ 
\cline{2-3}\cline{3-4}\cline{4-5}\cline{5-6}\cline{6-7}\cline{7-8}\cline{8-9}
& IBCF & \textit{C. Reduction} & \textit{Weight PoF} & \textit{Filter PoF} & IBCF & \textit{C. Reduction} & \textit{Weight PoF} & \textit{Filter PoF}\\ 
\hline
$\lbrace 50,100 \rbrace$ & 0.2991455 & \cellcolor{blue!25} 0.3045573 & 0.3232279 & \cellcolor{blue!25} 0.1299230 & 0.3089337 & \cellcolor{blue!25} 0.3128119 & \cellcolor{blue!25} 0.3314886 & \cellcolor{blue!25} 0.1300847 \\
\hline 
$\lbrace 15,20 \rbrace$ & 0.2991455 & \pmb{0.3447523} & \pmb{0.3808720} & 0.3499418 & 0.3089337 & \pmb{0.3538197} & \pmb{0.3893160} & 0.3515182 \\
\hline 
$\lbrace 2,7 \rbrace$ & 0.2991455 & 0.3151632 & 0.3676126 & \pmb{0.4129423} & 0.3089337 & 0.3255337 & 0.3755165 & \pmb{0.4138143} \\
\hline
\multicolumn{9}{|c|}{MAP results for $\alpha = 1$ ($BOW=0\%$ - $EN=70\%$ - $DT=30\%$)}\\
\hline
\multirow{2}{*}{Granularities} & \multicolumn{4}{c|}{$MAP@5$} & \multicolumn{4}{c|}{$MAP@10$}\\ 
\cline{2-3}\cline{3-4}\cline{4-5}\cline{5-6}\cline{6-7}\cline{7-8}\cline{8-9}
& IBCF & \textit{C. Reduction} & \textit{Weight PoF} & \textit{Filter PoF} & IBCF & \textit{C. Reduction} & \textit{Weight PoF} & \textit{Filter PoF}\\ 
\hline
$\lbrace 50,100 \rbrace$ & 0.2991455 & \cellcolor{blue!25} 0.3042942 & 0.3212988 & \cellcolor{blue!25} 0.1327955 & 0.3089337 & \cellcolor{blue!25} 0.3117196 & 0.3287964 & \cellcolor{blue!25} 0.1333247 \\
\hline 
$\lbrace 15,20 \rbrace$ & 0.2991455 & \pmb{0.3672521} & \pmb{0.4010373} & \cellcolor{blue!25} 0.3059922 & 0.3089337 & \pmb{0.3743193} & \pmb{0.4067024} & \cellcolor{blue!25} 0.3074584 \\
\hline 
$\lbrace 2,7 \rbrace$ & 0.2991455 & 0.3218124 & 0.3807921 & \pmb{0.4232245} & 0.3089337 & 0.3315916 & 0.3882532 & \pmb{0.4245341} \\
\hline
\multicolumn{9}{|c|}{MAP results for $\alpha = 1$ ($BOW=0\%$ - $EN=50\%$ - $DT=50\%$)}\\
\hline
\multirow{2}{*}{Granularities} & \multicolumn{4}{c|}{$MAP@5$} & \multicolumn{4}{c|}{$MAP@10$}\\ 
\cline{2-3}\cline{3-4}\cline{4-5}\cline{5-6}\cline{6-7}\cline{7-8}\cline{8-9}
& IBCF & \textit{C. Reduction} & \textit{Weight PoF} & \textit{Filter PoF} & IBCF & \textit{C. Reduction} & \textit{Weight PoF} & \textit{Filter PoF}\\ 
\hline
$\lbrace 50,100 \rbrace$ & 0.2991455 & 0.3295935 & 0.3444711 & \cellcolor{blue!25} 0.1806840 & 0.3089337 & 0.3375443 & 0.3518203 & \cellcolor{blue!25} 0.1821174 \\
\hline 
$\lbrace 15,20 \rbrace$ & 0.2991455 & \pmb{0.3918618} & \pmb{0.4245879} & 0.3416817 & 0.3089337 & \pmb{0.4004201} & \pmb{0.4325431} & 0.3437888 \\
\hline 
$\lbrace 2,7 \rbrace$ & 0.2991455 & 0.3151337 & 0.3627965 & \pmb{0.3907416} & 0.3089337 & 0.3239142 & 0.3696845 & \pmb{0.3918403} \\
\hline

\end{tabular}
\end{table*}

The presented results are for the three context-aware recommendation algorithms (\textit{C. Reduction}, \textit{Weight PoF} and \textit{Filter PoF}), and also for the baseline IBCF. The \textit{DaVI-BEST} results are not presented because they are equivalent to the IBCF results. So, there is no improvement
by using this algorithm and the contextual information extracted with our proposal.


The analysis of the results can be divided into 3 questions: 1) What is the algorithm with the best results?; 2) What is the granularity with the best results?; and 3) What is the value of combination factor with the best results?

Answering the first question, we can observe that the algorithm \textit{Weight PoF} presented the best results. This algorithm was also better than the baseline with statistical significance in all the experiments. For the second question, each algorithm was better at different granularity levels. The \textit{C. Reduction} and \textit{Weight PoF} algorithms presented the best results at configuration $\lbrace 15,20 \rbrace$, while the \textit{Filter PoF} algorithm presented the best results at configuration  $\lbrace 2,7 \rbrace$ (topics more specifics). The topics extracted by the configuration $\lbrace 50,100 \rbrace$ presented values of MAP not as high as for the others configurations, which shows that it is better to consider more specific topics and in larger amount.

Analyzing the values of combination factor, to answer the third question, \textit{C. Reduction} and \textit{Weight PoF} presented the best results for combination factor $\alpha = 0.3$, ($BOW = 70\%$ - $EN = 20\%$ - $DT = 10\%$) and ($BOW = 70\%$, $EN = 10\%$, $DT = 20\%$) respectively, and \textit{Filter PoF} for combination factor $\alpha = 0.5$ ($BOW = 50\%$ - $EN = 25\%$ - $DT = 25\%$).
      
\begin{figure}[!htb] \centering
\centering{\includegraphics[width=8cm,height=4.8cm]{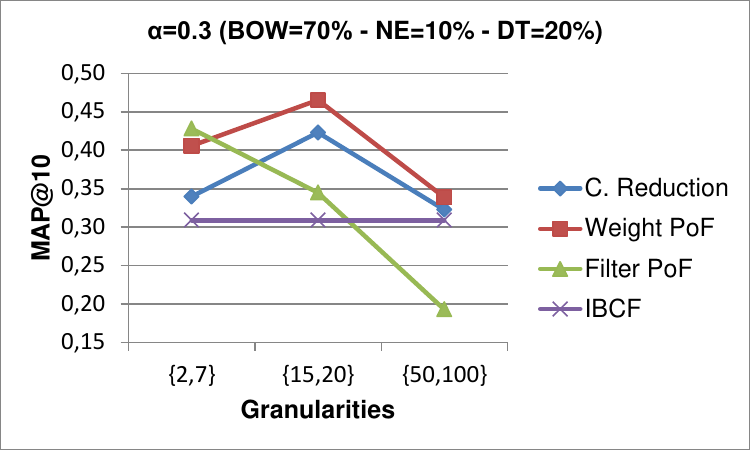}}
\caption{Graphic with values of MAP@10 for $\alpha=0.3$ ($BOW=70\%$, $EN=10\%$ and $DT=20\%$)} \label{fig:graphic}\end{figure}

In the graphic of Figure~\ref{fig:graphic}, we can analyze the best results of our experiments, i.e., the results for $\alpha=0.3$ ($BOW=70\%$, $EN=10\%$ and $DT=20\%$). The x-axis represents the granularities levels while the y-axis represents the values of MAP@10. Each line of the graphic is a recommender algorithm. It is evident that the three context-aware algorithms presented better results than the baseline \textit{IBCF}, only \textit{Filter PoF} presented a lower value of MAP at the granularity $\lbrace 50,100 \rbrace$. At the granularity $\lbrace 2,7 \rbrace$, this same algorithm presented better results than the others algorithms, what shows that this algorithm presents high values of MAP when topics more specifics are used. The algorithms \textit{Weight PoF} and \textit{Filter PoF} presented the best values of MAP at the granularity $\lbrace 15,20 \rbrace$, being the \textit{Weight PoF} the best of them.

\section{Conclusion}
\label{sec:conclusion}

In this paper, we proposed to use contextual information from topic hierarchies, constructed by LIHC method, to improve the accuracy of context-aware recommender systems. The topic hierarchies was constructed considering traditional bag-of-words (technical information), and the combination of named entities (privileged information I) and domains terms (privileged information II). The empirical evaluation showed that by using topics from the topic hierarchies with combined privileged information as contextual information, context-aware recommender systems can provide better recommendations. The contextual information obtained from the three topic hierarchies improved the recommendations in 3 out of 4 recommender systems evaluated in this paper: \textit{C. Reduction}, \textit{Weight PoF} and \textit{Filter PoF} (in most of the experiments).

As future work, we will finish some experiments in which we are comparing the combined use of the two types of privileged information against the results of our previous studies using named entities and domain terms separately. Additionally, we will also compare our proposal against other baselines proposed in the literature.

\section*{Acknowledgment}

The authors would like to thank FAPESP (grant \#2010/20564-8, \#2012/13830-9, and \#2013/16039-3, S\~ao Paulo Research Foundation (FAPESP)) and CAPES/Brazil for financial support.

\bibliographystyle{abbrv}
\bibliography{referencias}

\begin{thebibliography}{10}

\bibitem{Aciar2010}
S.~Aciar.
\newblock Mining context information from consumer's reviews.
\newblock In {\em CARS'10: Proceedings of the 2nd Workshop on Context-Aware
  Recommender Systems}, 2010.

\bibitem{Adomavicius2005b}
G.~Adomavicius, R.~Sankaranarayanan, S.~Sen, and A.~Tuzhilin.
\newblock {Incorporating Contextual Information in Recommender Systems Using a
  Multidimensional Approach}.
\newblock {\em ACM Trans. Inf. Syst.}, 23(1):103--145, 2005.

\bibitem{Adomavicius2011}
G.~Adomavicius and A.~Tuzhilin.
\newblock {Context-Aware Recommender Systems}.
\newblock In {\em Recommender Systems Handbook}, pages 217--253. Springer,
  2011.

\bibitem{Bauman2014}
K.~Bauman and A.~Tuzhilin.
\newblock {Discovering Contextual Information from User Reviews for
  Recommendation Purposes}.
\newblock In {\em CBRecSys '14: Proceedings of Workshop on New Trends in
  Content-based Recommender Systems}, pages 2--9, 2014.

\bibitem{Breese1998}
J.~S. Breese, D.~Heckerman, and C.~Kadie.
\newblock {Empirical Analysis of Predictive Algorithms for Collaborative
  Filtering}.
\newblock In {\em Proc. of the Fourteenth Conference on Uncertainty in
  Artificial Intelligence}, pages 43--52, 1998.

\bibitem{Cabre2001}
M.~T. Cabr{\'e} and R.~E.~J. Vivaldi.
\newblock Automatic term detection: a reviw of current systems.
\newblock In D.~Bourigault, C.~Jacquemin, and M.-C. L'Homme, editors, {\em
  Advances in Computational Terminology}, pages 53--88. John Benjamins,
  Amsterdam / Philadelphia, 2001.

\bibitem{Cardoso2008}
N.~Cardoso.
\newblock {REMBRANDT - Reconhecimento de Entidades Mencionadas Baseado em
  Rela��es e AN�lise Detalhada do Texto}.
\newblock In {\em Desafios na avalia��o conjunta do reconhecimento de
  entidades mencionadas: O Segundo HAREM}, pages 195--211. Linguateca, 2008.

\bibitem{Conrado2014}
M.~S. Conrado.
\newblock {\em Extra\c{c}{\~a}o autom{\'a}tica de termos simples baseada em
  aprendizado de m{\'a}quina}.
\newblock PhD thesis, Instituto de Ci{\^e}ncias Matem{\'a}ticas e de
  Computa\c{c}{\~a}o - Universidade de S{\~a}o Paulo - USP, 2014.

\bibitem{Conrado2013}
M.~S. Conrado, T.~A.~S. Pardo, and S.~O. Rezende.
\newblock {A Machine Learning Approach to Automatic Term Extraction using a
  Rich Feature Set}.
\newblock In {\em NAACL-HLT-SRW '13: Proceedings of the Conference of the North
  American Chapter of the Association for Computational Linguistics: Human
  Language Technologies - Student Research Workshop}, pages 16--23, Atlanta,
  Georgia, June 2013. Association for Computational Linguistics.

\bibitem{Deshpande2004}
M.~Deshpande and G.~Karypis.
\newblock {Item-based top-N Recommendation Algorithms}.
\newblock {\em ACM Trans. Inf. Syst.}, 22(1):143--177, 2004.

\bibitem{Dey2001}
A.~K. Dey.
\newblock {Understanding and Using Context}.
\newblock {\em Personal and Ubiquitous Computing}, 5(1):4--7, 2001.

\bibitem{Domingues2013a}
M.~A. Domingues, A.~M. Jorge, and C.~Soares.
\newblock {Dimensions As Virtual Items: Improving the Predictive Ability of
  top-N Recommender Systems}.
\newblock {\em Inf. Process. Manage.}, 49(3):698--720, 2013.

\bibitem{Domingues2014}
M.~A. Domingues, M.~G. Manzato, R.~M. Marcacini, C.~V. Sundermann, and S.~O.
  Rezende.
\newblock Using contextual information from topic hierarchies to improve
  context-aware recommender systems.
\newblock In {\em ICPR '14: Proceedings of the 22nd International Conference on
  Pattern Recognition}, pages 3606--3611, Aug 2014.

\bibitem{Hariri2011}
N.~Hariri, B.~Mobasher, R.~Burke, and Y.~Zheng.
\newblock {Context-Aware Recommendation Based on Review Mining}.
\newblock In {\em Proc. of the 9th Workshop On Intelligent Techniques for Web
  Personalization and Recommender Systems}, pages 30--36, 2011.

\bibitem{Ho2012}
S.-S. Ho, M.~Lieberman, P.~Wang, and H.~Samet.
\newblock {Mining Future Spatiotemporal Events and Their Sentiment from Online
  News Articles for Location-aware Recommendation System}.
\newblock In {\em Proc. of the First ACM SIGSPATIAL International Workshop on
  Mobile Geographic Information Systems}, pages 25--32, 2012.

\bibitem{Li2010}
Y.~Li, J.~Nie, and Y.~Zhang.
\newblock {Contextual Recommendation Based on Text Mining}.
\newblock In {\em Proc. of the 23rd International Conference on Computational
  Linguistics: Posters}, pages 692--700, 2010.

\bibitem{Marcacini2013}
R.~Marcacini and S.~O. Rezende.
\newblock {Incremental Hierarchical Text Clustering with Privileged
  Information}.
\newblock In {\em Proc. of the 2013 ACM Symposium on Document Engineering},
  2013.

\bibitem{Marcacini2010a}
R.~M. Marcacini and S.~O. Rezende.
\newblock Torch: a tool for building topic hierarchies from growing text
  collections.
\newblock In {\em IX Workshop on Tools and Applications. In 8th Brazilian
  Symposium on Multimedia and the Web}, 2010.

\bibitem{Mikheev1999}
A.~Mikheev, M.~Moens, and C.~Grover.
\newblock {Named Entity Recognition without Gazetteers}.
\newblock In {\em Proc. of the Ninth Conference on European Chapter of the
  Association for Computational Linguistics}, pages 1--8, 1999.

\bibitem{Panniello2012}
U.~Panniello and M.~Gorgoglione.
\newblock {Incorporating Context into Recommender Systems: An Empirical
  Comparison of Context-based Approaches}.
\newblock {\em Eletronic Commerce Research}, 12(1):1--30, 2012.

\bibitem{Ricci2011}
F.~Ricci, L.~Rokach, B.~Shapira, and P.~B. Kantor, editors.
\newblock {\em {Recommender System Handbook}}.
\newblock Springer, 2011.

\bibitem{Salton1983}
G.~Salton and M.~J. McGill.
\newblock {\em {Introduction to Modern Information Retrieval}}.
\newblock McGraw-Hill, Inc., 1983.

\bibitem{Sekine2004}
S.~Sekine.
\newblock {Named Entity: History and Future}.
\newblock 2004.

\bibitem{Sundermann2014}
C.~V. Sundermann, M.~A. Domingues, R.~M. Marcacini, and S.~O. Rezende.
\newblock Using topic hierarchies with privileged information to improve
  context-aware recommender systems.
\newblock In {\em BRACIS '14: Proceedings of Brazilian Conference on
  Intelligent Systems}, pages 61--66, Oct 2014.

\bibitem{Vapnik2009}
V.~Vapnik and A.~Vashist.
\newblock {A new learning paradigm: Learning using privileged information}.
\newblock {\em Neural Networks}, 22:544 -- 557, 2009.

\end{thebibliography}




\end{document}